\documentclass[letterpaper,journal]{IEEEtran}
\usepackage{amsmath,amsfonts}
\usepackage{cite}
\usepackage{mathtools}
\usepackage{algorithmic}
\usepackage{array}
\usepackage[caption=false,font=normalsize,labelfont=sf,textfont=sf]{subfig}
\usepackage{textcomp}
\usepackage{stfloats}
\usepackage{url}
\usepackage{verbatim}
\usepackage{tabularx}
\usepackage{booktabs}
\usepackage{multirow}
\usepackage{graphicx}
\usepackage{comment}
\usepackage{placeins}
\usepackage{xcolor}
\def\BibTeX{{\rm B\kern-.05em{\sc i\kern-.025em b}\kern-.08em
    T\kern-.1667em\lower.7ex\hbox{E}\kern-.125emX}}
\usepackage{balance}
\begin{document}

\title{Empirical characterization of the Translational acoustic-RF communication channel}

\author{Ann Pamela Mathew, 
 V. V. Reddy, S. L. Srinivas
\thanks{ Ann Pamela Mathew and 
 V. V. Reddy is with the Department of Electronics and Communication Engineering, International Institute of Information Technology Bangalore, India.}
\thanks{S. L. Srinivas is with the Naval Science and Technological Laboratory (NSTL), DRDO, Visakhapatnam, India 
}
}


\markboth{}{}
{}

\maketitle

\begin{abstract}
Translational acoustic-radio frequency (TARF) communication paves the way for translating information from an underwater acoustic signal to the over-the-air (OTA) electromagnetic receiver through the medium interface. The study and characterization of the channel is essential for establishing a reliable communication link. Although channel modeling has been extensively studied for OTA and underwater channels, the amplitude characteristics of the TARF cross-medium channel have not been investigated in comparison with well-known distributions to date. In this work, we define the signal model incorporating the effects of the wavefront-water surface interactions. With the help of numerical and graphical methods, we then attempt to characterize the cross-medium channel with empirical data using existing models developed for OTA and underwater channels. We further evaluate channel linearity and time invariance empirically. Observations from these studies over multiple experiments are detailed with additional discussions that enable better channel characterization to develop reliable and consistent cross-medium TARF communication in challenging scenarios.

\end{abstract}


\section{Introduction}
Ubiquitous communication between space, terrestrial, and underwater users is an important vision of sixth-generation (6G) standards. Although communication in respective mediums has found significant success, communication across mediums, particularly from underwater to air, is challenging due to the dissimilar characteristics of the two media. The mismatched electromagnetic, acoustic, and optical properties of the two media at their interface introduce reflection, refraction, and mode-conversion losses, thereby creating hard trade-offs in range, data-rate, alignment, latency, and robustness.

While acoustic, electromagnetic, and optical modes are employed for underwater communication \cite{theocharidis2025underwater,stojanovic2003acoustic}, physical translation and optical methods are applied for cross-medium communication \cite{qu2024review,luo2022recent,huang2021preliminary,cevik2015overview,carver2022air}. Among these approaches, translational acoustic-RF (TARF) communication \cite{tonolini2018networking,mollahosseini2024surf} has gained attention due to its potential to directly bridge underwater and over-the-air media exhibiting better robustness to misalignment between the transmitter and receiver. Furthermore, unlike optical communication, the underwater transmitter location is not compromised.

As the name suggests, the channel in TARF communication translates the information embedded in an underwater acoustic wave into the phase of the RF wave. This is accomplished by the use of a highly sensitive radar that can pick micron-level vibrations induced by the acoustic wave on the water surface. Despite such interaction between the two media, communication has been demonstrated experimentally in~\cite{tonolini2018networking}, where several modulation schemes are explored to maximize the data rate with high reliability.

The underwater acoustic wave propagating to the water surface induces a surface wave that propagates outwards radially~\cite{qu2021cross}. The acoustic information is translated into the micro-vibrations that ride on the surface wave. These micro-vibrations are embedded in the phase of the radar sensing the water surface from above. In literature\cite{tonolini2018networking,qu2021cross,mollahosseini2024surf,mollahosseini2024snooping}, this entire phenomenon is modelled as linear. However, the translation of information from one medium to the other compels an investigation of the channel for its linearity and time invariance in order to develop a reliable communication mechanism across the two media. 

In over-the-air (OTA) wireless systems, channel modeling is well established, with standardized statistical models such as Rayleigh and Rician fading widely used to represent non-line-of-sight and line-of-sight conditions, respectively \cite{rappaport2002wireless,goldsmith2005wireless}. These models form the foundation for theoretical analysis and system design across generations of wireless technologies \cite{proakis2001digital,proakis2001communication}. In contrast, underwater acoustic communication channels do not have a universally accepted model due to their strong dependence on environmental factors and their inherently time-varying and non-stationary nature. While certain statistical models may approximate channel behavior under specific and well-defined scenarios \cite{stojanovic2009underwater,ruizvega2011ricean}, underwater channel modeling generally relies on a combination of physics-based and scenario-dependent approaches, making it more complex and less standardized than OTA channel modeling \cite{pu2010underwater}. For cross-media communication where the information propagates through the water and air media, such channel modeling studies have not been performed so far.

In this work, we define the TARF communication channel incorporating the superposition of acoustic wave arrival at various points on the water surface. Subsequently, we conduct an empirical study on the cross-medium channel characteristics to examine its proximity to prominent distributions. We further verify the linearity and time-invariance characteristics from TARF communication experiments.


This paper is structured as follows. In Section $2$, we discuss the basic operation of the TARF communication system and define the signal model. The data set and the experimental setup are summarised in Section $3$. In Section $4$, we analyse the outcomes of the empirical evaluation. 


\section{Signal model for TARF communication}

In TARF communication, when the acoustic wave transmitted by the underwater source interacts with the water surface, the message is translated to the micro-vibrations that ride on the surface waves. These vibrations are sensed by an over-the-air frequency-modulated continuous wave (FMCW) radar. We here model the underwater wave propagation as well as the sensing of the micro-vibrations using the radar in more detail.

\begin{figure}[!ht]
        \centering
    \includegraphics[width=0.4\textwidth]
        {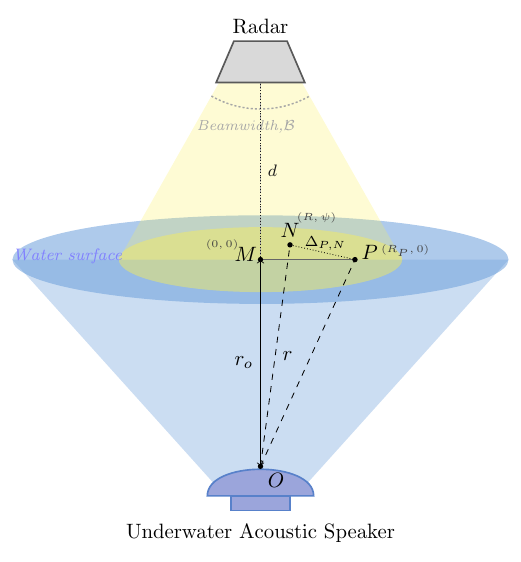}
        \vspace{-0.25cm}        
        \caption{ Underwater acoustic propagation and radar sensing in TARF communication.}
        \label{fig:TARF_SignalModel}
    \end{figure}

Consider an underwater acoustic emitter at a depth of $r_{o}$ as shown in Fig.~\ref{fig:TARF_SignalModel}.
\vspace{-0.5cm}
    \vspace{-0.15cm}
 The spherical underwater acoustic wavefront from the emitter approaches different points on the water surface at different time instances. We first denote the point on the water surface above the emitter as $M$ and consider the same as the origin of the water surface coordinate system. The pressure wave arriving at point $M$ is modeled as \cite {lurton2002introduction}
\begin{equation}
\begin{aligned}
p_i(\omega, t) &= A(r,\omega)\, e^{j\omega (t - r_{o}/c_{w})}\\&=\frac{A_0(\omega)}{r_{o}}\, e^{-\alpha_{1}r_{o}}\,e^{j\omega (t - r_{o}/c_{w})},
\end{aligned}
\end{equation}
where $A_0(\omega)$ \cite{tonolini2018networking}
represents the amplitude of the transmitted acoustic signal dependent on the angular frequency, $\omega$. $\alpha_{1}$ is the absorption coefficient and $c_{w}$ is the underwater sound speed.

When the acoustic pressure interacts with the surface, a damped surface wave originates and propagates radially outwards.  At any point $P(R_{P},0)$,  the effective
induced wave due to the underwater acoustic wave interacting
at point $M$ and the surface wave propagation from $M$ to $P$ is modelled as~\cite{qu2021cross}
 
\begin{equation}
\begin{aligned}
\delta_{M,P}(\omega,t)
= &
\frac{2 p_i(\omega,t)}{\omega \rho c_{w}}\,
e^{-\alpha_{2}R_{p}}
\cos\!\left(
kR_{p}-\omega t
\right) \\
= &
\frac{2 A_0(\omega)}{\omega \rho c_{w} r_{o}}\,
e^{-(\alpha_{1}r_{o}+\alpha_{2}R_{p})}e^{j\omega(t-\frac{r_{o}}{c_{w}})}
\cos\!\left(
kR_{p} - \omega t
\right),
\end{aligned}
\end{equation} 
where $R_{P}$ is the radial distance from the point $M$ to $P$, $\rho$ is the water density and $k$ is the surface wave number. This water surface disturbance accounts for only the acoustic wave path $\overline{OM}$ and surface wave path $\overline{MP}$ at the point $P$. Since the underwater acoustic wave interacts with the surface at different points at different time instances,  each such interaction generates a surface wave. 

We first define a sector $\Omega=\{(R,\psi):0\le R\le R_{P},
\ \psi_{\min}\le\psi\le\psi_{\max}\}$ such that the induced surface waves at various points within the sector propagate to point $P$ and superimpose prominently. The effective induced wave at $P$ can be written as
\begin{equation}
\begin{aligned}
\delta_P(\omega,t) =
\iint_{\Omega}
\frac{2A_0(\omega)}{\omega \rho c_{w}r}
e^{-(\alpha_{1}r+\alpha_{2}\Delta_{P,N}(R,\psi)}
\\
\times e^{j\omega(t-\frac{r}{c_{w}})}\cos(k\Delta_{P,N}(R,\psi) - \omega t)\, dRd\psi,
\end{aligned}
\end{equation}
where the variable $r = \sqrt{r_o^2 + R^2}$ and $\Delta_{P,N}(R,\psi)=\sqrt{(R_{P}-\cos\psi)^2+R^{2}\sin\psi^2}$ is the distance between point $P(R_{P},0)$ and any point $N$ lying in the sector $\Omega$. It is evident that $\delta_P(\omega,t)$ has the acoustic message $A_0(\omega)$ that is scaled by the integral.

The millimeter wave radar looking towards the water surface has a beamwidth, $\mathcal{B}$, as illustrated in Fig.~\ref{fig:TARF_SignalModel}. The response for the $l$th transmit chirp would be received after a propagation delay $ \mathcal\tau_l= \frac{2}{c}\big[d + w(t_l) + m(t_l)]$ 
where $t_l$ is the slow-time variable, and $d$ corresponds to the distance of the water surface from the phase center of the radar.  $\mathcal\tau_l$ also incorporates the response due to the surface waves, $w(t_l)$, and $\
m(t_l)=\iint_{\mathcal{B}} \beta\,\delta_P(\omega,t_l)\,\mathrm{d}\mathrm{B}$, which is the response due to the acoustic micro-vibrations induced on the watersurface over the radar beamwidth $\mathcal{B}$. 

We emphasize that the distance between the water surface and the radar varies within the wide radar antenna beam, as observed in Fig.~\ref{fig:TARF_SignalModel}. Consequently, the micro-vibrations induced over a large water surface area will be embedded over multiple range bins, albeit with low intensity. The received signal at a specific range bin, $r_0$, across the slow time is given by 
\begin{align}
            z(t_l) &= Z(f_r, t_l)\big|_{f_r = f_{r_0}} 
            = \overline{\alpha} \, e^{\jmath 2\pi f_0 \frac{2}{c}\big(m(t_l) + w(t_l)+ \phi\big)},
            \label{eq:slow time signal}
\end{align}
where $Z(fr,t_l) = \mathcal{F}\{z(t,t_l)\}$ is the received signal after the Fourier transform along the fast time, $ f_{r_0}$ is the frequency corresponding to the range bin, $r_0$, and $\phi$ is the residual phase due to the quadratic terms. The acoustic data is retrieved by unwrapping the argument of $z(t_l)$ as
\begin{equation}
        y(t_l) = \mbox{unwrap}\big(\mbox{arg}(z(t_l)) \big)= \overline{m}{(t_{l})} + \overline{w}{(t_{l})}+\overline{\phi},
\end{equation}
where $\overline{m}(t_{l})$ and $\overline{w}(t_{l})$ are proportional to the micro-vibration response ${m}(t_{l})$  and surface wave response ${w}(t_{l})$ respectively and $\overline{\phi}$ is the  constant phase variation. To suppress out-of-band noise, $y(t_l)$ is passed through a band-pass filter.

 In TARF communication, where the transmitted data is an acoustic signal and the received radar signal is electromagnetic, extracting the phase and determining the micro-motions is crucial to the overall performance of the cross-medium communication link. The channel through which this information passes plays a significant role in the translation of acoustic pressure waves into RF phase modulations. In the next section, we explain the channel characteristics before discussing the experimental setup for TARF communication. 


\section{Evaluation of TARF Communication channel }
\subsection{Channel characteristics}
In general, a communication system can be described by an operator that maps an input signal $x(t)$ to an output signal $y(t)$, written in generic form as
$y(t) = \mathcal{T}\{x(\cdot)\}(t)$,
where $\mathcal{T}\{\cdot\}$ represents the channel and receiver effects. Among the many possible system classes, linear models are most commonly adopted in communication theory, where  the input--output relationship reduces to
$y(t) = (h * x)(t) = \int_{-\infty}^{\infty} h(\tau)\,x(t-\tau)\,d\tau$,
where $h(t)$ is the channel impulse response.

In both over‑the‑air and underwater acoustic communication systems, the transmitted signal propagates through complex and often unpredictable physical environments. As a result, the received signal is affected by attenuation, multipath propagation, motion‑induced Doppler effects, and ambient noise. The amplitude of complex channel response is therefore modeled using various distributions, unlike an idealized scenario (static or purely LOS path) where $h(t)$ is a single deterministic component representing attenuation and phase shift. 

Over several decades, extensive studies have led to the development of multiple statistical channel models for OTA and underwater communication using Rayleigh, Rician, or Nakagami‑m distributions, to mention a few. However, such a study to model the entire cross-medium communication channel comprising not only the over-the-air and underwater channels but also the interface between the two has not been observed in the literature so far. In order to develop a robust and reliable communication system, one needs a more accurate and unified understanding of real‑time propagation phenomena across cross-medium communication systems. Hence, in the following sections, we conduct an empirical study of the cross-medium channel to verify if it follows any well-known distributions.

\subsection{Experimental Setup and data construction}\label{SCM}

\begin{figure}[ht]
        \centering
        \includegraphics[width=\linewidth]{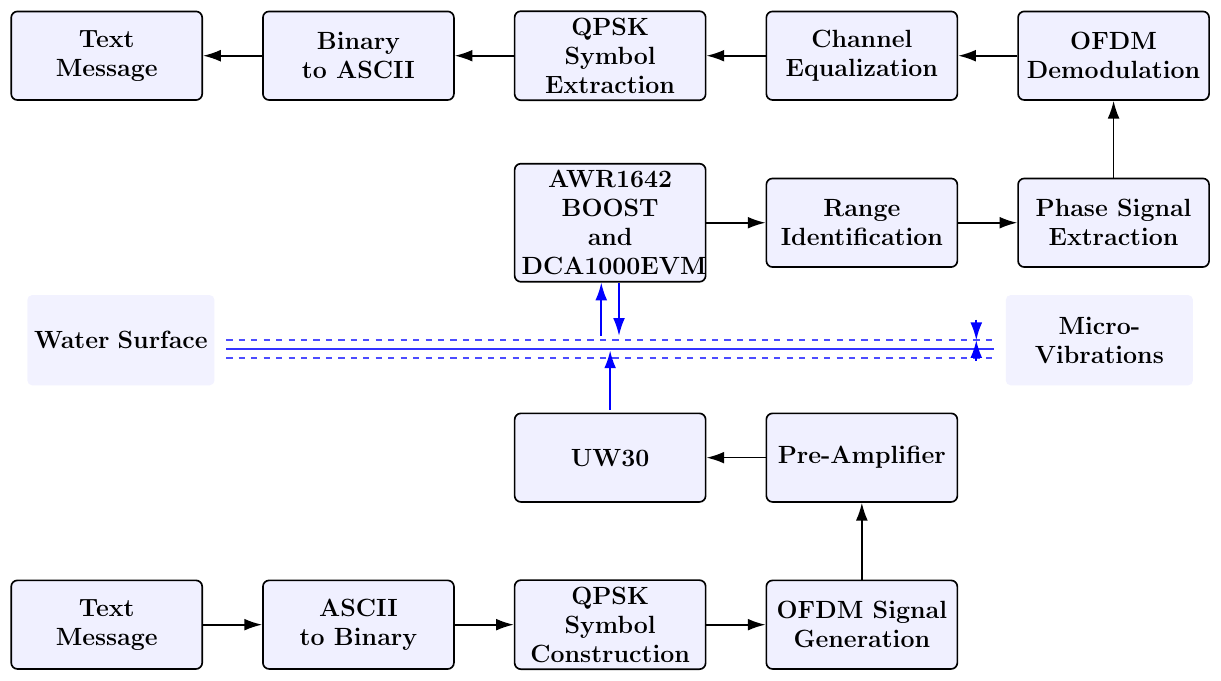}
        \caption{ End-to-end TARF communication system.}
        \label{fig:End_to_end_TARF_system}
    \end{figure}

To evaluate channel characteristics and their effects on the performance of the TARF communication system, we performed several experiments in a pool of dimensions $37~$m × $135~$m with a depth of $5~$m. The commercial underwater acoustic loudspeaker, Electro-Voice UW30, was employed to transmit the acoustic signals. The acoustic message is transmitted by the speaker through a Bluetooth-enabled power amplifier. The millimeter-wave FMCW radar, Texas Instruments' AWR1642BOOST, is mounted on a carriage about $0.5~$m above the water surface in alignment with the underwater speaker to extract the micro-vibrations. The data acquisition card, DCA1000EVM, is used to acquire the raw ADC output. Fig. \ref{fig:End_to_end_TARF_system} illustrates the various stages involved in the implementation of an end-to-end TARF communication system.

 Restricting the communication to only text messages, American Standard Code for Information Interchange (ASCII) characters are converted to a binary sequence before quadrature phase shift keying (QPSK) symbols are constructed with every pair of binary data. To construct an orthogonal frequency-division multiplexing (OFDM) symbol that is employed for acoustic communication~\cite{proakis2008digital,zhou2014ofdm}, $32$ orthogonal subcarriers within a bandwidth of $100~$Hz from $100-200~$Hz are used. Of the $32$ subcarriers, $16$ subcarriers carry known pilot messages while the remaining carry the QPSK data. A conjugate symmetric OFDM spectrum is constructed before the inverse Fourier transform to obtain a real OFDM signal that can be transmitted through the speaker. The final transmitted signal for each experimental trial consists of 15 OFDM symbols of duration $1~$s, preceded by the cyclic prefix of duration $0.5~$s. 
 This transmitted signal travels through the water to reach the surface, where it induces micro-vibrations. 
 
\FloatBarrier 

 The FMCW radar, facing the water surface, sequentially transmits chirps and simultaneously receives the response. Once the received radar signal is down-converted, the discrete Fourier transform (DFT) is performed on the fast-time chirp response to obtain the dominant range frequency that corresponds to the water surface. This complex DFT coefficient at the specific range across chirps yields the slow-time signal given in Eq.\eqref{eq:slow time signal}. The micro-vibrations are extracted from the phase of the slow-time signal after unwrapping to retrieve the transmitted OFDM-modulated signal. This signal then undergoes OFDM demodulation and channel equalization before QPSK symbols are extracted. The binary sequence thus obtained is used to construct the ASCII characters and therefore the text message as shown in Fig.~\ref{fig:End_to_end_TARF_system}.

\section{Results and Discussions}

\begin{figure}[t]
\centering


\includegraphics[width=\linewidth]{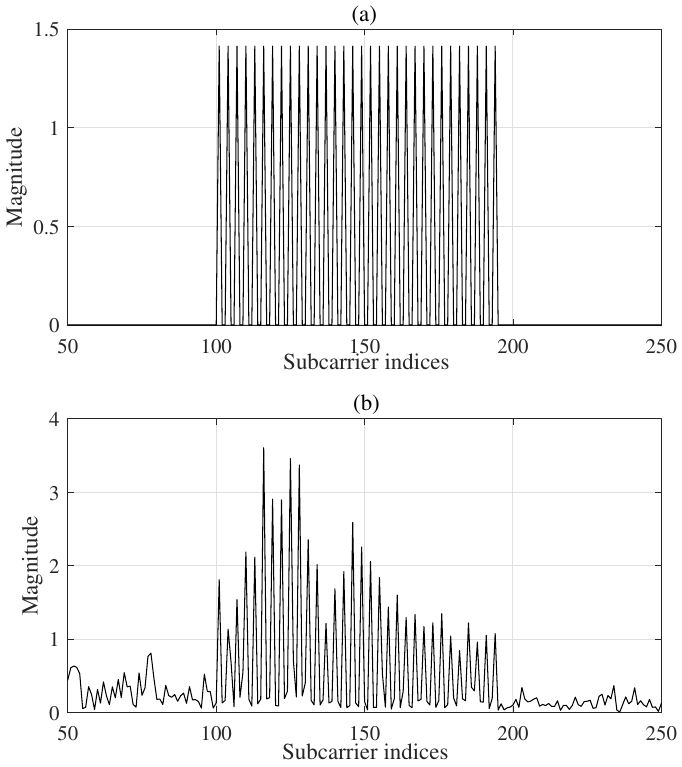}
\caption{Frequency Response of symbols for TARF: (a) Transmitted OFDM symbol (b) Received OFDM symbol}
\label{fig:Tx&Rx_OFDMsymbol}
\end{figure}

 As part of TARF communication channel analysis, it is first essential to verify that the micro-vibrations are induced on the water surface and the radar is sensitive to pick up these micro-vibrations. To affirm the same through an experiment, we transmit an OFDM signal with $32$ subcarriers, as shown in Fig.~\ref{fig:Tx&Rx_OFDMsymbol}(a). The received radar slow-time phase signal, $y(t_l)$, is expected to have this information at these subcarrier frequencies. 

The frequency spectrum of the received phase signal, plotted in Fig.~\ref{fig:Tx&Rx_OFDMsymbol}(b), shows a prominent response in all the transmitted subcarrier frequencies. However, unlike the transmitted signal spectrum, the spectrum of the received signal does not undergo uniform attenuation across subcarriers, indicating that the channel is frequency sensitive with severe attenuation towards the higher end of the spectrum. 

\subsection{Statistical analysis}

Noting the variations in the received subcarrier amplitudes, it is essential to characterize the statistics of the channel at these subcarriers with well-known channel distributions employed in OTA and underwater communications. Towards that, we employ the goodness-of-fit (GoF) techniques \cite{dagostino2017gof} to determine how well the empirical data distribution fits the corresponding theoretical distribution. In the GoF analysis, characteristics of the empirical data are compared with the corresponding quantities of the model, and the differences between them are assessed both numerically and graphically. 

Amongst the numerical GoF techniques, the Kolmogorov–Smirnov (KS) and Anderson-Darling (AD) tests are carried out to compare the empirical CDF of the observed data with the theoretical CDF of five well-known statistical models, viz., Rayleigh, Rician, Nakagami-m, Weibull, and  Rician-shadow. The KS test measures the maximum absolute deviation between the empirical and theoretical CDFs, thereby assessing overall distributional agreement. The AD test is similar to the KS test with special emphasis on the tails of the distribution. Furthermore, the overall root mean square error (RMSE) and tail-weighted RMSE between empirical and theoretical quantiles are also presented for all five distributions in Table~\ref{tab:gof_metrics} for speaker depths of $50~$cm and $70~$cm.

\begin{table}[t]
\centering
\caption{Goodness-of-Fit Statistics for various distributions}
\label{tab:gof_metrics}

\setlength{\tabcolsep}{4pt}
\footnotesize
\begin{tabularx}{\columnwidth}{l *{5}{>{\centering\arraybackslash}X}}
\toprule
\multirow[t]{2}{*}{GoF}
 & Rayleigh & Rician & Nakagami & Rician & Weibull \\
Metric\rule{0pt}{2.6ex}
 &  &  & $m$ & shadowed & \\
\midrule
\multicolumn{6}{c}{\textbf{Scenario 1: Depth=50cm}} \\
\midrule
KS
 & 0.105 & 0.105 & 0.073 & 0.082 & 0.133 \\
AD
 & 44.34 & 44.34 & 22.73 & 26.54 & 66.52 \\
\midrule
\multicolumn{6}{c}{} \\[-6pt]   
Overall
 &  &  &  &  &  \\
RMSE
 & 0.129 & 0.129 & 0.105 & 0.118 & 0.321\\
\multicolumn{6}{c}{} \\[-6pt]   
Tail-weighted
 &  &  &  &  &   \\
RMSE
 & 0.165 & 0.165 & 0.115 & 0.149 & 0.51  \\
\midrule
\multicolumn{6}{c}{\textbf{Scenario 2: Depth=70cm}} \\
\midrule
KS
 & 0.17 & 0.17 & 0.093 & 0.092 & 0.29 \\
AD
 & 111.8 & 111.8 & 29.4 & 28.4 & 319.8 \\
\midrule
\multicolumn{6}{c}{} \\[-6pt]   
Overall
 &  &  &  &  & \\
RMSE
 & 0.09 & 0.09 & 0.05 & 0.11 & 1.30 \\
\multicolumn{6}{c}{} \\[-6pt]   
Tail-weighted
 &  &  &  &  & \\
RMSE
 & 0.12 & 0.12 & 0.06 & 0.13 & 1.48\\
  &  &  &  &  & \\
\multicolumn{6}{@{}p{\linewidth}@{}}{%
\footnotesize\emph{Note:} Lower values indicate better agreement between empirical data and the theoretical distribution.
} \\
\bottomrule
\end{tabularx}
\end{table}

  Nakagami‑m and Rician‑shadowed distributions exhibit the smallest KS distances, indicating good global agreement across both depths. Rayleigh and Rician distributions show moderate agreement but larger deviations than Nakagami‑m distributions, while Weibull underperforms in comparison to other distributions. The AD tests confirm that the Nakagami‑m and Rician‑shadowed distributions best capture deep fading events. Rayleigh and Rician underperform in modeling extremes despite reasonable global fits. The overall and tail-weighted RMSE values reaffirm that Nakagami-m offers a balanced fit over the entire distribution compared to all other distributions.

  We next graphically compare the empirical CDF of the observed data with the theoretical CDF of the same five statistical models by constructing the probability–probability (P–P) plot \cite{mahmoud2021gof}.  A perfect alignment of the empirical CDF with the theoretical CDF yields a $45^\circ$ line indicating a good fit, while a systematic deviation from this line reveals a mismatch between the model and the data. 

 \begin{figure*}[t]
    \centering
        \centering
        \includegraphics[width=2\columnwidth]{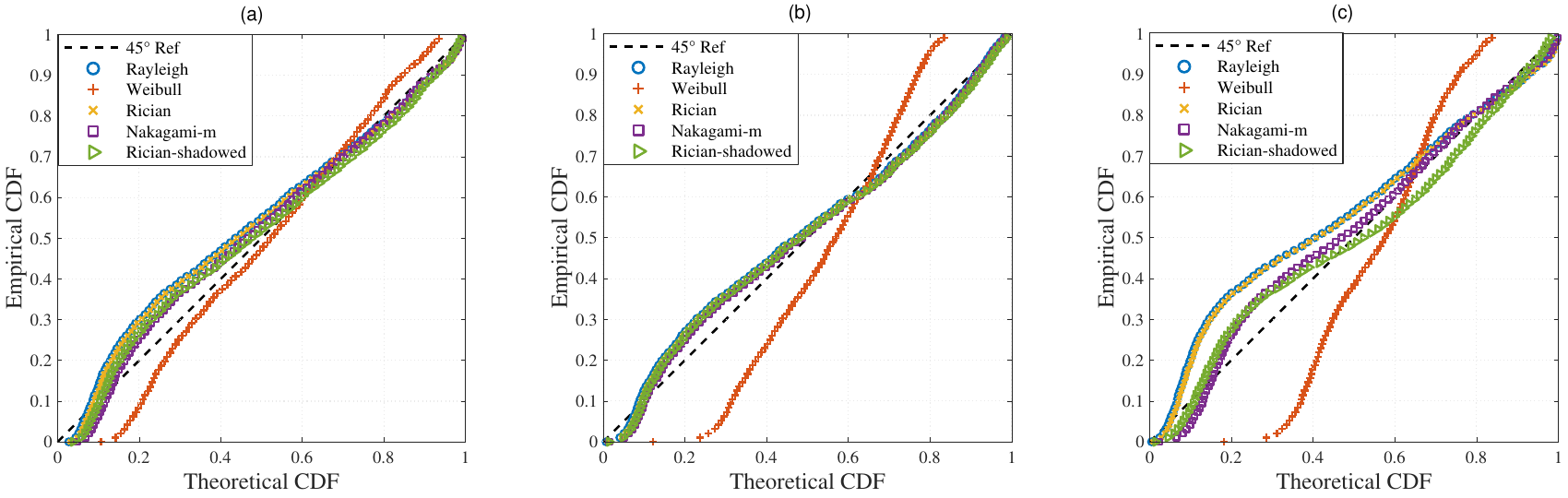}
        \vspace{-2mm}
    \caption{Combined probability--probability (P--P) plots for three experimental scenarios.(a) Scenario 1: Depth = 50 cm (b) Scenario 2: Depth = 60 cm (c)Scenario 3: Depth = 70 cm .}
    \label{fig:pp_combined}
\end{figure*}

  Fig.~\ref{fig:pp_combined} shows the P-P plot constructed from the empirical data for three different speaker depths. It can be observed from all three plots that the Nakagami-m model is closest to the $45^\circ$  reference line in comparison to the other models. However, this alignment with Nakagami-m distribution is mostly in the second half of the CDF range, while significant deviation is observed in the first half. It is also noted across the three subplots that these deviations vary with the depth of the speaker. In Fig.~\ref{fig:pp_combined}(a), the deviation between the two CDFs is observed only in the lower CDF range. This trend is observed in Figs.~\ref{fig:pp_combined}(b) and \ref{fig:pp_combined}(c) to change as depth increases. 

Although numerical GoF metrics in Table~\ref{tab:gof_metrics} assert that Nakagami-m and Rician-shadowed distributions model the channel better than the other distributions, the P-P plots reveal their considerable mismatch with the true underlying data distribution. Hence, the true channel statistics are likely more complex than any single classical parametric distribution. In view of these observations and the presence of the nonlinear unwrapping function in the phase signal retrieval, we next empirically evaluate the channel's fundamental properties of linearity and time invariance.
   
 \subsection{Evaluation for Linearity and Time invariance}
 \subsubsection{Linearity }
\begin{figure}[t]
    \centering

        \centering
        \includegraphics[width=\columnwidth]{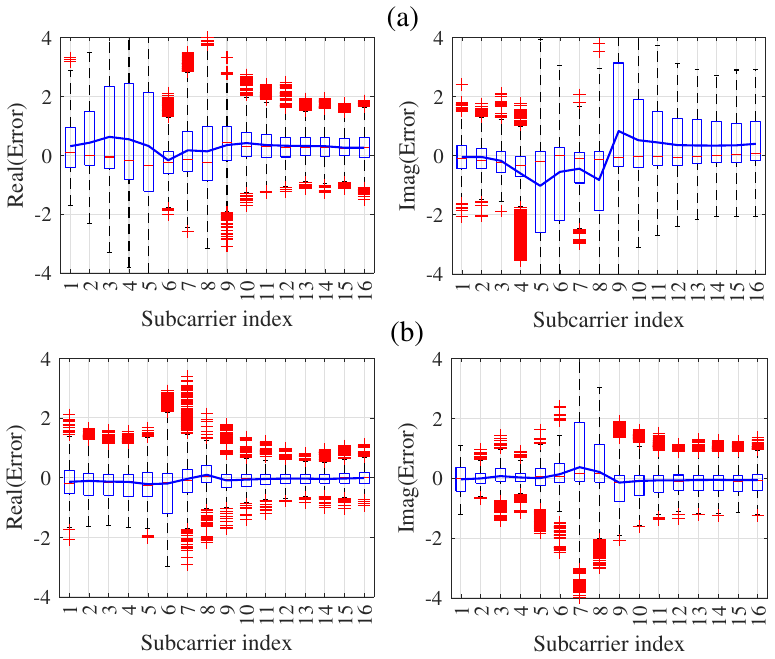}
        \vspace{-3mm}
        \caption{Box plot analysis for residual error, ${e}_{\text{lin}}$ for speaker depths (a) $50$ cm  (b)$70$ cm. Here, the blue solid line is the mean and the + are the outliers.}
    \label{fig:BoxPlotErrors}
\end{figure}

Given $X_{ip}$, $X_{jp}$, and $X_{kp}$ as the pilots for the $p^{\text{th}}$ sub carrier frequency for any $i$, $j$ and $k^{\text{th}}$ experiments, we verify linearity by constructing 
\begin{equation}
\begin{aligned}
X_{kp} &= a X_{ip} + b  X_{jp} .\\
\end{aligned}
\end{equation}
Since each pilot is modulated using the same QPSK symbol in our experiments, the weights to obtain $X_{kp}$ are given by $a=b=0.5$. We verify the channel linearity by checking for 
\begin{equation}
\begin{aligned}
Y_{kp}
&= a\,Y_{ip} + b\,Y_{jp},
\end{aligned}
\end{equation}
where $Y_{kp}=\mathcal{T}\{X_{kp}\}$, $Y_{ip}=\mathcal{T}\{X_{ip}\}$ and $Y_{jp}=\mathcal{T}\{X_{jp}\}$, are the pilots extracted from the received signal. If Eq.(8) is not satisfied, it indicates deviation from linearity. This condition is empirically studied by observing the residual error,
$\mathrm{e}_{\text{lin}}(k,p)$ 
 as given below.
\begin{equation}
\mathrm{e}_{\text{lin}}(k,p)= Y_{kp} - \left( a\,Y_{ip} + b\,Y_{jp} \right).
\end{equation}
Fig.~\ref{fig:BoxPlotErrors} presents the mean (blue line), median (red dash), and the variance (box size) of the residual error evaluated over all the experiments, along with the outliers (red markers) across all the subcarriers for speaker depths of $50$ and $70$ cm, respectively. We observe the non-zero bias and error variance (box size) to be inconsistent from subcarrier to subcarrier, with a varying number of outliers.

It can also be noted that some subcarriers show visibly skewed distributions (longer positive or negative tails). This asymmetry in the box plots about zero indicates that the channel is evolving.

\subsubsection{Time invariance}
To investigate the time invariance behavior of the channel, we transmit $K$ OFDM symbols one after the other in the same acquisition with identical pilots. To observe the evolution of the channel, the deviation in the received pilot response is studied over these symbols with respect to the first symbol, $Y_{ref}$, as the reference. The normalized error to verify time variance is defined as
\begin{equation}
\mathrm{e}_{\text{tiv}}(k)= \sum_{p}\left| \frac{Y_{ref}(p)}{\|Y_{ref}(p)\|} - \frac{Y_{k+1}(p)}{\|Y_{k+1}(p)\|} \right|^2,
\end{equation}
where $Y_{ref}(p)$ and $Y_k(p)$ are the $p^{\text{th}}$ sub-carrier pilots and $k=[1,K-1]$ is the relative symbol index. For a time-invariant system, $e_{tiv}$ remains consistent over symbols, while this error varies with $k$ for a time-variant channel.

\begin{figure}[t]
        \centering
        \includegraphics[width=\columnwidth]{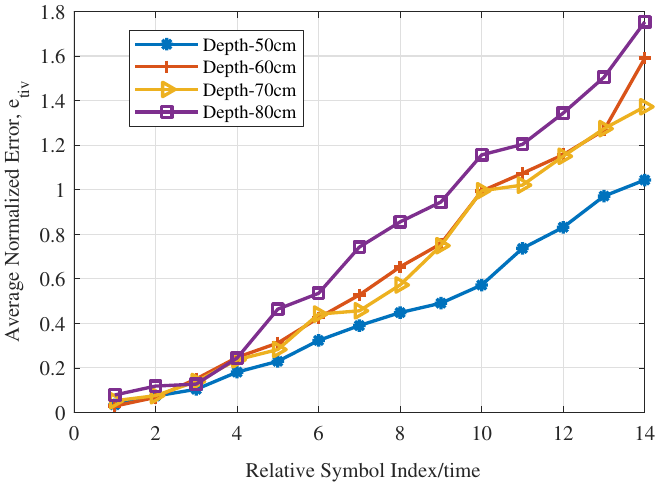}
        \caption{Residual TIV Error for End-to-End TARF communication system.}
        \label{fig:TIV error End_to_end_TARF_system}
    \end{figure}

Fig.~\ref{fig:TIV error End_to_end_TARF_system} presents the average normalized residual error, $\mathrm{err}_{\text{tiv}}(k)$,  against the relative symbol index w.r.t. the reference symbol for speaker depths varied from $50$cm to $80$cm.  It is observed that the residual error increases with time as well as the speaker depth. 

The increase in error observed with symbol index/time can be attributed to varying surface dynamics and/or the insufficient time in between the OFDM symbols for the micro-vibrations from the previous symbol to settle down~\cite{alazard2023damping}. The decay rate of waves/micro-vibrations on the surface is also sensitive to the medium viscosity~\cite{armaroli2018viscous,shelton2025time}.

To summarize the experimental observations, the TARF channel is first observed to deviate from common channel models employed in underwater and OTA. This is mainly due to the interface between the two media within the channel and the translation of the information from acoustic to EM regime. We next verified the entire TARF channel for its linearity empirically by comparing the transmitted and received pilots across all the OFDM subcarriers. The residual error statistics exhibit non-zero bias and fluctuating variance across subcarriers, indicating channel behavior that deviates from linearity. One factor that contributes to this is the unwrapping function that enables the extraction of micro-vibrations.

Finally, the channel is observed empirically to be time varying due to the surface dynamics and the absence of settling time between the symbol transmissions. As shown in the literature, surface waves (micro-vibrations) have a finite settling time depending on the medium properties\cite{mollahosseini2024snooping} such as viscosity, surface tension, and density that can impact communication over time, thereby requiring channel adaptation.

\section{Conclusion}
The study and characterization of the channel is essential for establishing a reliable communication link. In this work, we delve into these aspects for TARF communication, where information propagates across multiple media. We defined the signal model taking into consideration the effect of wavefront interactions at multiple points on the water surface. We then specifically focused on characterizing the cross-medium channel using well-known OTA and underwater channel models. To accomplish that, numerical and graphical methods were employed to compare theoretical distributions with empirical data acquisitions. Observations revealed that empirical data do not align with any one classical model. This study further emphasizes that the channel properties are time-varying and deviate from linearity. Potential reasons for such observations are identified to be cross-medium interface, translation from acoustic to EM regime, and medium-dependent decay rate for micro-vibrations. The analysis also highlights the need for advanced
channel adaptation mechanisms to ensure more dependable
and steady cross-medium communication links.

\section*{Acknowledgment}
This work is supported by the Naval Science and Technological Laboratory (NSTL), Vishakapatnam, India. 

\bibliographystyle{myIEEEtran}
\bibliography{ref}

\end{document}